# Magnetoresistance peculiarities and magnetization of materials with two kinds of superconducting inclusions

O. N. Shevtsova

National Technical University of Ukraine "Kiev Polytechnic Institute", 37 Prospect Peremogy, Kyiv, 03056, Ukraine

E-mail: oksana.shevtsova@kpi.ua

**Abstract.** Low-temperature properties of a crystal containing superconducting inclusions of two different materials have been studied. In the approximation that the inclusions' size is much smaller than the coherence length/penetration depth of the magnetic field the theory for magnetoresistance of a crystal containing spherical superconducting inclusions of two different materials has been developed, and magnetization of crystals has been calculated.

## 1. Introduction.

One of the results of the rapid development of nanotechnology is creation of various types of composite materials or structurally heterogeneous systems, which consist of a matrix (host material) and disperse inclusions, and are characterized by the properties that are absent in the material components. Depending on the shape and size of these inclusions such composite materials have different properties.

The contact doping was initially proposed as an alternative method for production of alloys from non-mixing components, which is different from traditional alloying or sintering technologies. The new technology is based on anomalously quick migration of components in the systems with monotectic transformation [1]. The proposed solution has made it possible to remove all limitations regarding chemical composition, microstructure uniformity and volume of the final products - the limitations which are inherent to sintering and alloying of composites, and thus allowed one to create new composite materials, which production has been considered impossible before that.

The contact doping technology allows to get Al-Cu-Pb alloys, containing up to 20% of Cu and 30% of Pb with uniform distribution of Pb in the alloy volume in the form of spherical inclusions encapsulated into intermetallic shell, Cu-Pb-Bi and Cu-Pb-Sn alloys in which inclusions of heavy low-melt elements are uniformly spread in the Cu matrix [1]. The problem of micro-structural irregularities and uncontrolled dispersion was solved by transmission of electric current pulses of definite duration, amplitude and shape through a sample.

Modern technologies actively use also the method of dispersed fillers injection to modify material properties, such as increased strength and service life, and to reduce the production cost of a new structural material, just by changing the type of inclusions. Technology-controlled structures or ordered composites are of special interest. Examples of such structures are indium-opal composites formed by the pressure-induced injection of indium into periodically located submicron pores of opal dielectric matrix. The resulting composite with a lattice of indium granules is characterized by 2-step run of temperature-resistance plot and the size dependence of the critical temperature and critical magnetic field [2 - 8].

Grain sizes in the composite materials vary from a few nanometres to several hundred nanometres, and the materials themselves are characterized by rather specific and unusual electric properties, such as electron tunnelling and Josephson links between the grains in superconducting state [9].

The formation of structurally inhomogeneous systems is not only a technologically controlled process. In the multi component systems it is observed an effect of segregation and creation of microscale inclusions of other phases, such as precipitation of metal phase [10]. Nuclear irradiation or doping of complex compounds such as semiconductors leads to creation of a structurally inhomogeneous material which is characterized by new properties.

One of the methods to create a new phase is a method of ion implantation, and the phases created by it are called the "ion beam synthesized phases" [11]. Among the recent applications of this method is the synthesis of superconducting nanocrystals of $MgB_2$. The presence of ion-implanted nanostructures can completely change the physical and chemical properties of a crystal. The latest achievement of the ion implantation technology is its role in creation of surface superconductivity in single crystals of $SrFe_2As_2$ [12]. Experiments with the magnetization and resistance of single crystals irradiated by ions $K^+$ and $Ca^{2+}$ (at a certain dose of irradiation) showed that there is a superconducting phase transition with the temperature slightly below 25 K. The surface superconductivity occurs in a layer, which is determined by the penetration depth of the ions.

An important aspect of the composite systems study is the use of their known properties and characteristics to identify the structural composition of new structurally inhomogeneous material formed as a result of irradiation or doping. Semiconductors of III - V groups, a typical representative of which is indium arsenide, are also the complex structures in which precipitation, i.e. the loss of another phase was observed [13]. It is known that precipitation of such crystal phase may be caused by a variety of technological processes, such as the dissociation of solid solutions [14]. If the metallic phase is formed, then by cooling a sample to a certain temperature we can get a crystal with superconducting inclusions in it. Superconductivity under high pressure in non-doped semiconductors GaSb, GaAs and GaP has been identified long time ago [15]. The phenomenon of superconductivity was discovered also in many chemical elements, alloys and in doped semiconductors. The conductivity features, which can be interpreted as a phase transition to the superconducting state, were found in the binary semiconductors PbTe [16 - 19]. The appearance of superconductivity in GaAs with deviations from normal stoichiometric composition was also observed in [20].

The search for new materials for spintronics led to intensive research of doped semiconductors [21 - 23]. The unexpected result of these studies was the discovery of superconducting gallium precipitates and chromium precipitates in the bulk samples of GaAs and GaP, alloyed with chromium. Magnetic measurements confirmed that the critical parameters of gallium ($T_c \approx 6,2$ K i $H_c \approx 600$ E) is the characteristic ones for type I superconductors [24].

Presence of superconducting inclusions leads to a jump of the sample's conductivity at low temperatures and to strong dependence of conductivity on the magnetic field (magnetoresistance). The occurrence of jump-like behaviour of magnetoresistance caused by the phase transition of inclusions from superconducting to normal state with increase of magnetic field was explained in the framework of the theory of magnetoresistance of crystals containing superconducting inclusions [25-27]. Specific features of magnetoresistance observed in InAs, irradiated by $\alpha$-particles [29], also indicate the presence of a phase transition. Since in this case the energy of the particles is very high (80 MeV), the indium-enriched metallic regions can be created in the crystal as the result of exposure, and at low temperatures they may become superconducting. In the framework of the magnetoresistance model of the crystal with randomly placed superconducting inclusions the calculations of the magnetoresistance of irradiated crystals were performed for different values of temperature, and the calculation results were compared with the available experimental data. Peculiarities of magnetoresistance observed in experiments were qualitatively explained in the framework of the magnetoresistance theory [30-32].

It should be noted that the calculation of magnetoresistance of complex materials is an important method to detect the presence of inclusions in multi-component samples.

An important method for detecting impurities in complex compounds is also plotting of dependencies of magnetization on magnetic field because low-temperature features of magnetization detected in the experiment under certain temperature indicate the presence of non-uniform inclusions. This method also allows estimating the size of inclusions and calculating their concentration.

## 2. The conductivity of the crystal with two types of superconducting inclusions

Let's calculate the conductivity of a system containing superconducting spherical inclusion that randomly located in the crystal. We believe that the total amount of inclusions or concentration of impurities is not sufficient for occurrence of superconductivity in the whole sample, i.e., the system is below the percolation threshold. Since the formation of the metallic phase is not technologically controlled, it would be logical to assume that the formed superconducting inclusions are characterized by dispersion of a certain size. In calculating the conductivity it can be assumed that, depending on the temperature and magnetic field, an inclusion can exist in two states: in superconducting state with infinite conductivity or in normal state, characterized by resistance, corresponding to the inclusion of material at a certain temperature.

The theory of magnetoresistance of a crystal with superconducting inclusions [24 - 27] is based on the assumption that the concentration of superconducting regions is low, the amount of inclusions in order of magnitude coincides with the coherence length and the critical magnetic field of the I type superconducting

inclusions is described by the well-known Ginzburg formula [32]:

$$H_c^{inc} / H_c = \sqrt{20}\frac{\lambda}{R}, \qquad (1)$$

where $\lambda = \lambda(T, T_c)$ is the magnetic field penetration depth; $H_c$ is the critical field of a bulk superconductor; $R$ is an inclusion radius. That is, in the framework of this theory the structure of the superconducting inclusion is not taken into account, and is considered homogeneous. In the case when the size of the superconducting spherical inclusions is larger than the coherence length / penetration depth of the magnetic field, it is necessary to take into consideration the vortex structures which are to be born in such superconducting inclusions.

The conductivity of the system depends on the volume of superconducting inclusions and matrix conductivity. To calculate the conductivity of the system the method of effective medium is used [33]. Let's calculate the conductivity of a crystal containing two types of spherical superconducting inclusions; such inclusions are generally characterized by different critical temperatures and varying dispersion. We'll use the formula for the conductivity of multi component systems [34, 35]

$$P_{1s}\frac{\sigma-\sigma_{1s}}{\sigma+2\sigma_{1s}} + P_{1n}\frac{\sigma-\sigma_{1n}}{\sigma+2\sigma_{1n}} + P_{2s}\frac{\sigma-\sigma_{2s}}{\sigma+2\sigma_{2s}} + P_{2n}\frac{\sigma-\sigma_{2n}}{\sigma+2\sigma_{2n}} + (1-P_1-P_2)\frac{\sigma-\sigma_h}{\sigma+2\sigma_h} = 0 \qquad (2)$$

where $\sigma_{is} = \infty$ is the conductivity of $i$-type of inclusions in the superconducting state, $\sigma_{in}$ is the conductivity of $i$-type of inclusions in the normal state, $\sigma_h$ is the conductivity of a matrix, $P_{is}$ and $P_{in}$ is the relative amount of inclusions in the superconducting and normal states, respectively, index $i = 1,2$ corresponds to inclusions of type I and II, respectively

$$P_{is} = P_i \frac{\int_0^{R_c^i(T,H)} R^3\, W_i(R)\, dR}{\int_0^\infty R^3\, W_i(R)\, dR}, \quad P_{in} = P_i - P_{is}, \qquad (3)$$

where $P_i$ is the relative volume of inclusions of $i$-type $P = P_1 + P_2$ is the full relative amount of inclusions in the sample, $W_i(R)$ is the probability that in the unit interval with the radius R one can found the inclusion of $i$-type. For numerical calculations we have used a normal distribution of inclusions by their radius with dispersion $s_i$ and radius $R_{0i}$

$$W_i(R) = Z\exp\left[-\frac{(R-R_{0i})^2}{2s_i^2}\right], \qquad (4)$$

where Z is determined from the normalization condition $\int_0^\infty W_i(R)dR = 1$.

It should be noted that the lower limit of integration in equation (3) is to be determined by some minimal radius of an inclusion, defined as the limit value, which allows one to use the Ginsburg-Landau approximation. And as the size distribution of the inclusions is chosen in such a way that the amount of very small inclusions (and therefore their contribution to the conductivity) is negligibly low, so, conventionally, the minimum radius can be considered as zero.

To calculate the effective conductivity $\sigma$ of the crystal containing two types of superconducting inclusions, which are generally characterized by two different critical temperatures $T_{c1}$ and $T_{c2}$, two different values of Ginzburg-Landau parameters $\kappa_1$ and $\kappa_2$, and by different dispersion, is necessary to solve equation (2). The effective conductivity of such a system is the value that is determined by many parameters: the relative volume of inclusions, the average size of inclusions and material properties of superconducting inclusions. Key parameters of the system are the temperature and the external magnetic field, because by changing them one can induce the phase transition of the system from superconducting to normal state and thus adjust the relative volume of superconducting (normal) inclusions. Therefore we'll consider the temperature dependence of conductivity for different values of the magnetic field and specific features of the magnetic resistance at fixed values of temperature.

## 3. Temperature dependence of conductivity

Let consider the system containing two types of inclusions. The critical temperature of the inclusions of type I is lower than the critical temperature of inclusions of type II, i.e. $T_{c1} < T_{c2}$ For the calculations a dielectric matrix was considered which contains Sn and Pb inclusions with critical temperatures $T_{c1}^{Sn}$ = 3.7 K, and $T_{c2}^{Pb}$ = 7.2 K, respectively. Then the dynamics of the phase transition of inclusions caused by temperature changes, should be considered for three cases: 1) $H = 0$; 2) $H < (H_c^{(1)}, H_c^{(2)})$; 3) $H < H_c^{(2)}$, where $H_c^i$ is the characteristic value of the critical field for inclusions of $i$-type. The results of the temperature dependence of conductivity for the corresponding value of the magnetic field for inclusions with different dispersion values are presented in figure 1. Since the matrix contains two different types of superconducting inclusions, a double-jump of the conductivity is observed at low temperatures. In absence of an external field ($H = 0$) the jumps are very sharp (see figure 1, curve 1), because in this model the critical temperature in absence of the magnetic field does not depend on the radius, so the phase transition is realized simultaneously for all inclusions with the same temperature. In the applied magnetic field, the phase transition of superconducting inclusions depends on the radius of the inclusions, and therefore at a given temperature $T$ only the inclusions with $R \leq R_{ci}(T, H, T_{ci})$ are in the superconducting state. Accordingly, at ($H < (H_c^{(1)}, H_c^{(2)})$) a smeared 2-step phase transition (see figure 1, curves 2a, 2b, 3a, 3b) is observed in the system, and the temperature region, which is characterized by high conductivity, decreases with the increase of magnetic field. The result of the further increase of the external field (situation $H < H_c^{(2)}$) (see figure 1, curves 4a, and 4b) is the disappearance of superconductivity in the inclusions of type I, and the conductivity of the system is characterized by the smeared single-step dependence, which is caused by the phase transition of type II inclusions. It is clear that the degree of smearing of the phase transition is determined by the dispersion value.

The dynamics of inclusions transition from the superconducting into normal state is illustrated in figure 2 for a fixed dispersion value ($s = 0,01$) and for different values of the magnetic field. The relative volume of the inclusions is 1% and 2 % for inclusions of type I and II, respectively. It is seen that at the minimum value of the field ($H / H_c^{(2)} = 0.16$) one can observe phase transitions for inclusions of I and II types and the phase transition is sharp (curves 2a and 2b); with further increase of the magnetic field phase transitions occur earlier (curves 3a and 3b), and the phase transition begins to smear, and at subsequent increase of the external field values only very smeared phase transition of type II inclusions can be observed (curves 4a and 4b). One can see that the relative amount of superconducting inclusions become lower with the increase of the magnetic field and, respectively, the relative volume of inclusions that have turned into the normal state, is increased. Computational parameters of the system: $r_0 / \xi_0 = 0,2$; $P_1 = 0,01$; $P_2 = 0,02$; $T_{c1} = 3,7K$; $T_{c2} = 7,2K$; $\kappa_1 = 0,13$; $\kappa_2 = 0,23$; $\sigma_1 / \sigma_h = 6$; $\sigma_2 / \sigma_h = 3$.

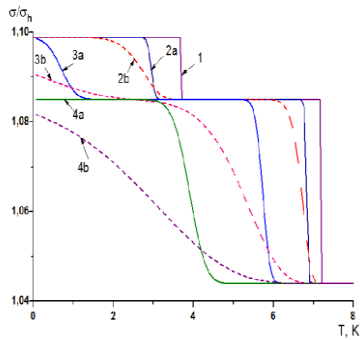 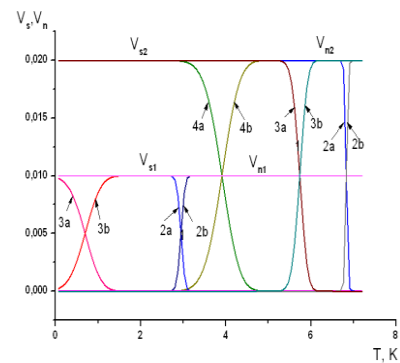

Figure 1. Temperature dependence of conductivity for different values of the magnetic field:
$H / H_c^{(2)} =$: 1) 0; 2) 0.16; 3) 0.3; 4) 0.5;
a) $s = 0,01$; b) $s = 0,02$.

Figure 2. Dynamics of superconducting and normal inclusions at $s = 0,01$ and the same values of the magnetic field. Letters 'a' and 'b' indicate the superconducting and normal state of inclusions, respectively.

## 4. Low Temperature Conductivity Peculiarities in Applied Magnetic field

Peculiarities of conductivity versus applied magnetic field should also be considered for 3 temperature ranges:
1) $T < T_{c1}$ - all inclusions of $R \leq (R_{c1}(T,H), R_{c2}(T,H))$ are in the superconducting state, 2) ($T_{c1} < T < T_{c2}$) only the type II inclusions remained in the superconducting state, 3) ($T > T_{c2}$) - all inclusions turned back into the normal state. The results of computation for conductivity as magnetic field function for inclusions with different dispersion are shown in figure 3. One can see that at ($T < T_{c1}$) there is a strong 2-step conductivity (magnetoresistance) (curves 1a, and 1b), which decreases with the increase of magnetic field. In the temperature range ($T_{c1} < T < T_{c2}$) the high conductivity area decreases (curves 2a, and 2b), and at $T > T_{c2}$ the phase transition is realized only for inclusions with a higher critical temperature (curves 3a, 3b).

Such peculiarity of magnetoresistance is caused by suppression of superconductivity first in the larger inclusions, and then, at increase of the magnetic field the smaller inclusions become involved. This phenomenon is shown in figure 4, which presents the dependence of conductivity on the magnetic field for different values of the average size of inclusions. The range of the magnetic field that is characterized by high conductivity, is the largest in the case of the smallest average size of inclusions (curve 1a), and with the increase of the average size of the inclusions (curves 1b, 1c) the areas with high conductivity become smaller. Growth of temperature also significantly reduces the area of high conductivity ($1a \rightarrow 2a$), ($1b \rightarrow 2b$), the range of magnetic fields in which magnetoresistance is decreased is determined by the average size of inclusions, and the area of such decrease is regulated by variance. Thus, the temperature and field dependencies of the conductivity are mainly determined by the size and variance of inclusions. The next computational parameters of the system were used: $r_0/\xi_0 = 0,2$; $P_1 = 0,01$; $P_2 = 0,02$; $T_{c1} = 3,7K$; $T_{c2} = 7,2K$; $\kappa_1 = 0,13$; $\kappa_2 = 0,23$; $\sigma_1/\sigma_h = 6$; $\sigma_2/\sigma_h = 3$.

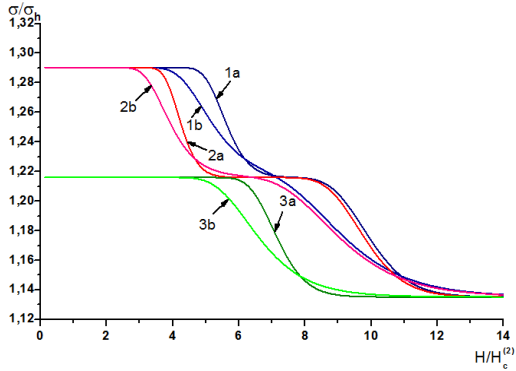
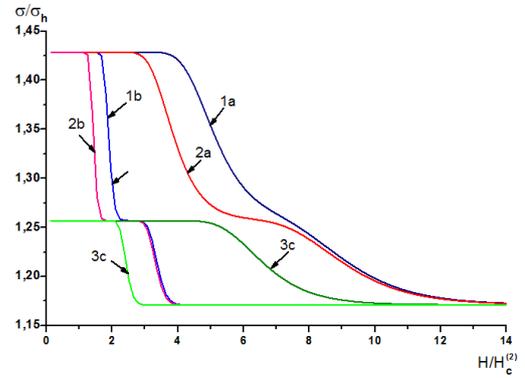

Figure 3. The conductivity of the system as a function of magnetic field at different values of temperature 1) $T = 1K$; 2) $T = 3K$; 3) $T = 6K$; a) $s = 0,01$; b) $s = 0,02$; $r_0/\xi_0 = 0,2$; $P_1 = 0,01$; $P_2 = 0,05$.

Figure 4. The dependence of the conductivity on the magnetic field for inclusions of different sizes: a) $r_0/\xi_0 = 0,1$; b) $r_0/\xi_0 = 0,2$; $s = 0,02$; 1) $T = 1K$; 2) $T = 3K$; 3) $T = 6K$. $P_1 = 0,05$; $P_2 = 0,05$.

## 5. Magnetization of a crystal with different kinds of superconducting inclusions

To calculate the effective magnetization of the crystal containing spherical inclusions of different types, it is necessary to determine the magnetization of an individual inclusion, and then perform the procedure of averaging the magnetization of individual inclusions, which takes into account the dispersion of inclusions on the radius, concentration of inclusions and their distribution in the host crystal.

To calculate the magnetization of a single superconducting inclusion is necessary to write the self-consistent system of GL equations with the relevant boundary conditions on the surface of the inclusions. And since we restrict our consideration by the inclusions of small radii, the length of which is less than the coherence length, than in this case an order parameter that characterizes the superconducting state can be considered constant, and only the second-order GL equation can be considered for the magnetic field, which in a spherical coordinate system with the beginning in the centre of the inclusion of radius $R$ can be written as:

$$\frac{1}{\rho^2}\frac{\partial^2}{\partial \rho^2}(\rho^2 A_{in}) + \frac{1}{\rho^2 \sin(\theta)}\frac{\partial}{\partial \theta}(\sin(\theta)\frac{\partial A_{in}}{\partial \theta}) = \frac{1}{\lambda^2}A_{in}, \qquad \text{at } \rho < R, \qquad (5)$$

$$\frac{1}{\rho^2}\frac{\partial^2}{\partial \rho^2}(\rho^2 A_{out}) + \frac{1}{\rho^2 \sin(\theta)}\frac{\partial}{\partial \theta}(\sin(\theta)\frac{\partial A_{out}}{\partial \theta}) = 0, \qquad \text{at } \rho > R. \qquad (6)$$

Solutions of equations (5) and (6) can be obtained by separation of variables

$$A_{in} = \sum_{n=1}^{\infty}\left(C_n\sqrt{\frac{\pi}{2}}I_{n+1/2}\left(\frac{\rho}{\lambda}\right)/\sqrt{\frac{\rho}{\lambda}} + D_n\sqrt{\frac{\pi}{2}}K_{n+1/2}\left(\frac{\rho}{\lambda}\right)/\sqrt{\frac{\rho}{\lambda}}\right)\frac{d(P_n(\cos(\theta)))}{d\theta}, \qquad (7)$$

where $I_{n+1/2}\left(\frac{\rho}{\lambda}\right)$, $K_{n+1/2}\left(\frac{\rho}{\lambda}\right)$ are modified Bessel functions; $P_n(\cos(\theta))$ is Legendre polynomial. Since the solution at zero must be finite, then $D_n = 0$. Similarly, we can write the solution of equation (6) for the vector potential outside the sphere:

$$A_{out} = \sum_{n=1}^{\infty}\left(a_n\rho^n + \frac{b_n}{\rho^{n+1}}\right)\frac{d(P_n(\cos(\theta)))}{d\theta}. \qquad (8)$$

The radial and angular components of the magnetic field $H_r$ and $H_\vartheta$ were found from the equation

$$\vec{H} = rot\vec{A}. \qquad (9)$$

From the condition of continuity of the radial and angular component of the magnetic field on the sphere surface the expressions for the distribution of the magnetic field in a spherical superconductor can be obtained:

$$H_r^{in} = -2\frac{C_1\lambda^2}{r^3}\left(-\frac{r}{\lambda}\cosh\left(\frac{r}{\lambda}\right) + \sinh\left(\frac{r}{\lambda}\right)\right)\cdot\cos(\theta), \qquad (10)$$

$$H_\theta^{in} = -\frac{C_1}{r^3}\lambda^2\left(-\frac{r}{\lambda}\cosh\left(\frac{r}{\lambda}\right) + \left(\frac{r}{\lambda}\right)^2\sinh\left(\frac{r}{\lambda}\right) + \sinh\left(\frac{r}{\lambda}\right)\right)\cdot\sin(\theta), \qquad (11)$$

$$H_r^{out} = H_0\cdot(1+\frac{2b_1}{\rho^3})\cdot\cos(\theta), \qquad (12)$$

$$H_\theta^{out} = -H_0\cdot(1+\frac{b_1}{\rho^3})\cdot\sin(\theta), \qquad (13)$$

where

$$C_1 = \frac{3}{2}\frac{H_0 R}{\sinh\left(\frac{R}{\lambda}\right)}, \quad b_1 = -\frac{1}{2}H_0 R^3 + \frac{3}{2}H_0 R^2 \frac{\cosh(\frac{R}{\lambda})}{\sinh\left(\frac{R}{\lambda}\right)} - \frac{3}{2}H_0\lambda^2 R. \qquad (14)$$

The value of the magnetization is calculated by the formula

$$-4\pi M = H_0 - H. \qquad (15)$$

If the size of the inclusion is small enough $R/\lambda \to 0$, then one can obtain the classic expression for the magnetic moment of a spherical superconducting inclusion [36]

$$m = -\frac{1}{30}H_0 R^3\left(\frac{R}{\lambda(T)}\right)^2. \qquad (16)$$

Let's calculate magnetization of the crystal containing superconducting inclusions of two types. In papers [24] and [18] experimental measurements were fulfilled of magnetization as a function of the magnetic field for doped semiconductors containing modified Ga-inclusions with (Tc=6.2K) and Pb-inclusions. Peculiarities of magnetoresistance were observed in the experiment [28] as well, and computation of the magnetoresistance [29-31] have shown that irradiation of a crystal creates radiation-induced spherical inclusions enriched with indium, which are characterized by a certain variance of sizes.

The calculation of magnetization versus magnetic field was fulfilled for a crystal containing inclusions of small sizes ($R \ll \lambda$). In this case, the magnetization behaviour is determined by the dynamics of the transition of superconducting inclusions of two different materials. For computation the Sn and Pb inclusions were considered, which are the type I superconductors, and are characterized by the following values of critical parameters: $T_{c1} = 3,7K$; $T_{c2} = 7,2K$; $\kappa_1 = 0,13$; $\kappa_2 = 0,23$. For simplification of our consideration we can assume that both types of inclusions are characterized by the same size and the same variance, but by different values of part of inclusions. In this case, the magnetization is characterized by two minima of different depth; each of them is caused by a specific type of inclusions. It can be seen (figure 5 and figure 6) that at temperature increasing one of the minima caused by phase transitions of inclusions of the I type disappears, and with further increase of temperature the magnetization value is decreased. Thus, the obtained dependence characterizes the presence of inclusions of various materials that are in the superconducting state, and the presence of two minima (or more in more complex samples) indicates the presence of appropriate number of types of inclusions in the material. That is, if the experimental results of magnetization of the material are characterized by such type of behaviour, than it can be stated that inclusions of some other material are incorporated in the crystal. Moreover, changing the temperature of a sample in the course of the experiment, we can determine rather accurately the exact type of material of the inclusions in the sample, their size and variance.

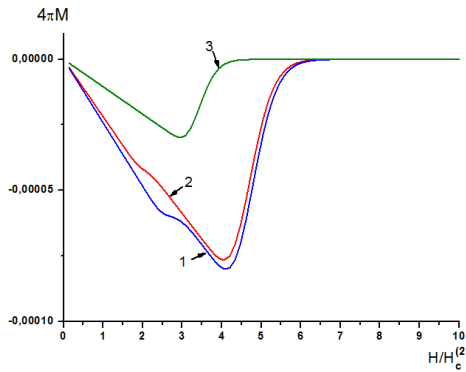
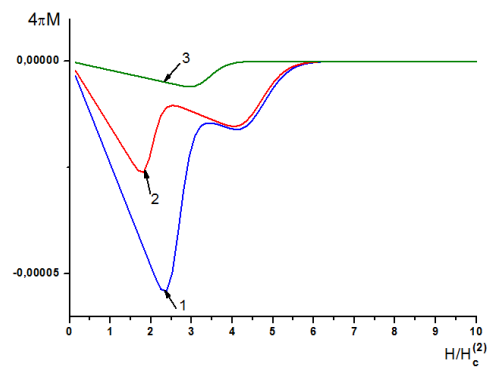

Figure 5. Magnetization versus magnetic field of the material containing Sn and Pb inclusions at different temperatures: 1) $T = 1K$ ; 2) $T = 3K$ ; 3) $T = 6K$. Parts of superconducting inclusions are equal to $P_1 = 0,01$, $P_2 = 0,05$. $H_c^{(2)}$ is the critical field of a Pb bulk sample.

Figure 6. Magnetization versus magnetic field of the material containing Sn and Pb inclusions at different temperatures: 1) $T = 1K$ ; 2 $T = 3K$ ; 3) $T = 6K$. $s = 0,01$ ; $r_0 / \xi_0 = 0,2$. Parts of superconducting inclusions are equal to $P_1 = 0,05$, $P_2 = 0,01$.

In the third temperature range ($T > T_{c1}$) the behaviour of the magnetization or diamagnetic response, caused by the presence of superconducting inclusions of one type in the crystal can be interpreted as a phase transition of only Pb superconducting inclusions with the change of the field (for fixed values of temperature) (figures 5,6, curves 3). It is seen that the appropriate magnetization curve consists of a linear and non-linear part, and with growth of temperature the magnetization value is decreased as well.

**6. Conclusions**

Thus, the presence of superconducting inclusions significantly changes physical properties of a crystal. The conductivity at low temperatures is increasing and there is a strong dependence of conductivity on the magnetic field, and the magnetic field range in which high conductivity is realized, increases with decreasing of the size of inclusions. This dependence is caused by phase transitions of inclusions from the superconducting to the normal state with the increase of magnetic field. The obtained results can be used for correct explanation of the conductivity at low temperatures in binary and more complex semiconductors, in which the precipitation of the superconducting phase is possible during the technological processing or under external impact. These characteristics of electrical conductivity and magnetic properties were observed in PbTe, PbJ$_2$, InAs, GaAs, GaP, where the metal- enriched phase precipitation is possible (the lead in PbTe and PbJ$_2$, indium in InAs, GaAs and gallium in GaP).

The presence of inclusions can be revealed at measurement of low-temperature magnetization, which is characterized by the characteristic minima caused by phase transitions of various types of superconducting inclusions in the magnetic field. Depths of these minima are determined by the volume and sizes of inclusions, and their variance.

**References**

1. Avraamov Yu.S., Shlyapin A.D. 2004 Mechanical engineering and engineering education. #1. 48-50.
2. Graf M. J., Huber T. E. and Huber C.A. 1992 *Phys. Rev. B* **45** 3133 - 3136.
3. Charnaya E.V., Tien C., Lin K. J. and Wur C. S. 1998 *Phys. Rev. B.* **58**, No. 1. 467 - 472.
4. Shamshur D.V., Chernyaev A.V., Fokin A.V. and Romanov S.G. 2005 *Physics of the Solid State* **47** 2005

- 2014.
5. Lungu A., Bleiweiss M, Amirzadeh J., Saygi S., Dimofte A., Yin M., Iqbal Z. And Datta T. 2001 *Physica C.* **349** 1 - 7.
6. Michotte S., Piraux L., Dubois S.,Pailloux F., Stenuit G. And Govaerts J. 2002 *Physica C.* **377** 267 - 276.
7. Aliev A.E., Lee S.B., Zakhidov A.A. and Baughman R.H. 2007 *Physica C.* **453** 15 - 23.
8. Butler J.F., Hartman T.C. J. Electrochem. 1969 *Soc.: Sol. St. Sci.* -. **116** 260 - 263.
9. Gadza M., Kuzz B., Murawskii L., Stizza S. and Natali R. 2008 *Optica Applicata* **XXXYIII** 153 - 161.
10. Russel K.C. 1984 *Prog. Mat. Sci.* **28**, No. 3. 229 - 434.
11. Kirkby K.J. and Webb R.P. 2003 Ion Implanted Nanostructures // *Encyclopedia of Nanoscience and Nanotechnology* / Ed. by H.S. Nalwa **4**. 1 - 11.
12. Chong S.V., Tallon J. L., Fang F., Kennedy J., Kadowaki K. And Williams G.V.M. 2011 *Europhys. Lett.* **94** 37009 p1- 37009- p6.
13. Buzea C. and Robbie K. 2005 *Supercond. Sci. Technol.* **18 - R1**
14. Milvidskii M.G. and Osvenskii V.B. 1984 *Structural defects in monocrystals of semiconductors.* Moscow: Metallurgy 256
15. Berman I.V., Brandt N. B. and Sidorov V. I. 1971 *Pisma. Zh. Eksp. Teor. Fiz* **14**, No. 1. 11 - 12.
16. Dedekaev T.T., Moshnikov V.A., Chesnokova D.B. and Yaskov D.A. 1980 *Pisma. Zh. Eksp. Teor. Fiz.* **6** 1030 - 1033.
17. Darchuk S.D., Ditl T., Korovina L.A., Kolesnik S., Savitskii M. and Sizov F.F. 1998 *Fizika I Technika polyprovodnikov*. **32** 786 - 789.
18. Darchuk L.A., Darchuk S.D., Sizov F.F. and Golenkov A.G. 2007 *Fizika i Technika polyprovodnikov*. **41** 144 - 148.
19. Takaoka S., Sugita T. and Murase R. 1987 *JJAP. Supplement.* **26**. 1345 - 1346.
20. Baranowski J.M., Liliental-Weber Z., Yau W.-F. and Weber E.R. 1991 *Phys. Rev. Lett.* **66**, 3079 - 3082.
21. Ohno H. 1999 *J. Magn. Magn. Mater.* **200** 110 - 129.
22. Dietl T. 2002 *Semicond. Sci. Technol.* **17** 377 - 392.
23. Pearton S. J., Abernathy C. R., Overberg M. E., Thaler G.T., Norton D.P., Theodoropoulou N., Hebard A.F., Park Y.D., Ren F., Kim J. and Boatner L.A. 2003 *J. Appl. Phys* **93** 1. 1 - 13.
24. Gosk J.B., Puzniak R., Strzelecka G., Materna A., Hruban A.,Wisniewski A., Szewczyk A, Kowalskii G., Korona K., Kaminska M. And Twardowskii A. 2008 *Supercond. Sci. Technol.* **21**. 065019 - 065024.
25. Sugakov V.I. and Shevtsova O.N. 2000 *Supercond. Sci. Technol.* **13** 1409 - 1414.
26. Sugakov V.I. and Shevtsova O.N., 2001 *Low Temp. Phys.* **27** 88 - 92.
*27.* Sugakov V.I. and Shevtsova O.N., 2001 *Proceedings of Institute for Nuclear Research* № 3(5). 100- 104.
28. Vikhlii G.A., Karpenko A.Ya. and Litovchenko P.G. 1998 *Ukrainian Physical Journal*. **43** 103- 106.
29. Sugakov V.I., Shevtsova O.N., Karpenko A.Ya., Litovchenko P.G. and Vikhlii G.A. 2003 *Low Temp. Phys.* **29** 551 - 555.
30. Mikhailovskii V.V., Sugakov V.I., Shevtsova O.N. Karpenko A.Ya., Litovchenko P.G. and Vikhlii G.A. 2007 *VANT. Physics of radiation damages and radiation study of materials*. **90** 55 - 62.
31. Ginzburg V.L. 1958 *JETP* **34** 113 - 125.
32. Kirkpatrick S**.** 1973 *Rev. Mod. Phys.* **45** 574 - 588.
33. Shklovskii B.I. and Efross A.L., *1975 Uspekhi physicheskich nauk* **117** 401-435.
34. Springett B.E. 1973 *Phys. Rev. Lett.* **31** 1463 - 1465.
35. Lifshits E.M. and Pitaevskii L.P., *1978 Statistical physics, part 2, Moscow: Nauka* – p.447